\documentclass[journal,twoside,web]{ieeecolor}
\usepackage{generic}
\usepackage{cite}
\usepackage{amsmath,amssymb,amsfonts}
\usepackage{algorithmic}
\usepackage{graphicx}
\usepackage{textcomp}
\usepackage{bm} 
\usepackage{multirow}
\usepackage{booktabs}
\usepackage{threeparttable}
\newcommand{\argmax}{\mathop{\rm arg\ max}\limits}
\def\BibTeX{{\rm B\kern-.05em{\sc i\kern-.025em b}\kern-.08em
    T\kern-.1667em\lower.7ex\hbox{E}\kern-.125emX}}
\begin{document}
\title{Non-Gaussianity Detection of EEG Signals Based on a Multivariate Scale Mixture Model for Diagnosis of Epileptic Seizures}
\author{Akira~Furui${}^{*}$,~\IEEEmembership{Member,~IEEE,}
        Ryota~Onishi,
        Akihito~Takeuchi,
        Tomoyuki~Akiyama${}^{*}$,
        and~Toshio~Tsuji${}^{*}$,~\IEEEmembership{Member,~IEEE}
\thanks{${}^{*}$A.~Furui, R.~Onishi, and ${}^{*}$T.~Tsuji are with the Graduate School of Engineering, Hiroshima University, Higashi-hiroshima, 739-8527 Japan (e-mail: akirafurui@hiroshima-u.ac.jp; tsuji@bsys.hiroshima-u.ac.jp).}
\thanks{A.~Takeuchi is with the Department of Neonatology, Okayama Medical Center, National Hospital Organization.}
\thanks{${}^{*}$T.~Akiyama is with the Department of Child Neurology, Okayama University Hospital (e-mail: takiyama@okayama-u.ac.jp).}
\thanks{Copyright (c) 2020 IEEE. Personal use of this material is per-mitted. However, permission to use this material for any other pur-poses must be obtained from the IEEE by sending an email to pubs-permissions@ieee.org. This paper is an accepted version for publication in IEEE Transactions on Biomedical Engineering.}}

\maketitle

\begin{abstract}
\textit{Objective:}
The detection of epileptic seizures from scalp electroencephalogram (EEG) signals can facilitate early diagnosis and treatment.
Previous studies suggested that the Gaussianity of EEG distributions changes depending on the presence or absence of seizures; however, no general EEG signal models can explain such changes in distributions within a unified scheme.
\textit{Methods:}
This paper describes the formulation of a stochastic EEG model based on a multivariate scale mixture distribution that can represent changes in non-Gaussianity caused by stochastic fluctuations in EEG.
In addition, we propose an EEG analysis method by combining the model with a filter bank and introduce a feature representing the non-Gaussianity latent in each EEG frequency band.
\textit{Results:}
We applied the proposed method to multichannel EEG data from twenty patients with focal epilepsy.
The results showed a significant increase in the proposed feature during epileptic seizures, particularly in the high-frequency band.
The feature calculated in the high-frequency band allowed highly accurate classification of seizure and non-seizure segments [area under the receiver operating characteristic curve (AUC) = 0.881] using only a simple threshold.
\textit{Conclusion:}
This paper proposed a multivariate scale mixture distribution-based stochastic EEG model capable of representing non-Gaussianity associated with epileptic seizures.
Experiments using simulated and real EEG data demonstrated the validity of the model and its applicability to epileptic seizure detection. 
\textit{Significance:}
The stochastic fluctuations of EEG quantified by the proposed model can help detect epileptic seizures with high accuracy.
\end{abstract}


\section{Introduction}
\IEEEPARstart{E}{pilepsy} is a chronic neurological disorder characterized by recurrent and unprovoked epileptic seizures caused by abnormal electric activities of brain neurons.
It is reported that there are approximately 50 million epilepsy patients worldwide~\cite{WHO}.
Electroencephalogram (EEG) can record neuronal electric activities and is routinely utilized for the diagnosis of epilepsy.
Especially prolonged EEG recording is a powerful tool to confirm an epileptic seizure when a patient manifests signs and/or symptoms that are difficult to judge from clinical observation or when there are no visible symptoms.
It is important to recognize seizures early because prolonged seizures can lead to increased brain damage.
However, the correct interpretation of EEG requires extensive training, and EEG specialists may not be readily available in emergency settings.
Tools that aid non-expert clinicians in interpreting EEG can eventually improve neurological outcomes through early recognition and treatment of seizures.

Numerous studies have attempted to detect epileptic seizures quantitatively using EEGs~\cite{Acir2005,Subasi2005,Nishad2020-jx,Alotaiby2015-bg,kellaway1979precise,Deburchgraeve2008,Greene2008,Sharma2020-ko, Sharma2020-rg,mooij2020skew,Srinivasan2007-yw,Ocak2009-rh,Arunkumar2018-cq}.
In these studies, EEGs are generally processed by two steps: frequency decomposition and feature extraction.
First, frequency decomposition divides EEGs into sub-bands to obtain frequency components that strongly reflects the effects of epileptic seizures.
Filter bank techniques based on wavelet transform~\cite{Subasi2005,Nishad2020-jx} and band-pass filters~\cite{Alotaiby2015-bg} have been widely used in this step.
In the feature extraction step, features that characterize the difference between seizures and non-seizures are then calculated from the EEG sub-bands.
Feature extraction plays an important role in detecting epileptic seizures, as the extracted features are generally used as input for machine learning-based classification schemes.
In many cases, simple amplitude information is used to characterize EEGs during epileptic seizures~\cite{kellaway1979precise,Deburchgraeve2008,Greene2008}.
However, detecting epileptic seizures by evaluating only the magnitude of the amplitude in specified sub-bands is sometimes difficult because EEGs recorded during epileptic seizures will differ significantly with the patient's age and epilepsy type.
Therefore, in recent years, the relationship between epileptic seizures and amplitude-independent features, such as higher-order statistics~\cite{Sharma2020-ko, Sharma2020-rg,mooij2020skew} and entropy~\cite{Srinivasan2007-yw,Ocak2009-rh,Arunkumar2018-cq}, has been studied by many researchers.

Meanwhile, the probability distribution shapes of EEGs have also been studied as a feature independent of amplitude~\cite{Saunders1963,Campbell1967,Weiss1973,Charles1999,Nurujjaman2009,Gonen2012}.
Previous studies typically assumed that EEG signals followed a steady Gaussian distribution~\cite{Saunders1963,Gonen2012}; however, after Campbell \textit{et al.} statistically indicated that EEG signals had non-Gaussian properties~\cite{Campbell1967}, various studies examined the probability distribution shapes of EEG signals, focusing on their Gaussianity~\cite{Weiss1973,Charles1999,Nurujjaman2009,Gonen2012}.
Although few previous studies investigated the relationship between epileptic seizures and the Gaussianity of EEG signals, they reported that EEG signals during epileptic seizures follow a non-Gaussian distribution~\cite{Charles1999,Nurujjaman2009}.
However, these studies are only experimental reports, and currently, there is no theoretical framework for quantitatively addressing the non-Gaussianity of EEG signals.
Building such a framework could lead to the development of a novel feature for detecting epileptic seizures.

In this study, we assume that non-Gaussianity in EEG signals is caused by stochastic fluctuations in amplitude, and propose a multivariate scale mixture model that can estimate these fluctuations from recorded EEG signals.
In the proposed model, a scalp EEG signal recorded at a certain time using multichannel electrodes follows a multivariate Gaussian distribution, and its variance-covariance matrix can be considered as a random variable following an inverse Wishart distribution.
Thereby, it allows the evaluation of the stochastic fluctuation of the variance-covariance matrix corresponding to the EEG amplitude.
In addition, by combining this model with a filter bank, we develop an analysis method that can calculate a feature representing time-series stochastic fluctuations in each EEG frequency band.
In the simulation experiment, parameter estimation based on the proposed method is conducted on artificial data, and the error between true values and estimated values observed when conditions are changed is discussed.
In the analysis experiment, the proposed method analyzes EEG signals recorded during focal seizures.
We also verify the goodness-of-fit of the proposed model and investigate the correspondence between (1) estimated stochastic fluctuations of EEG signals and (2) diagnoses by an epileptologist.

The reminder of this paper is organized as follows: Section II outlines the structure of the multivariate scale mixture model, the parameter estimation method, and the EEG analysis method considering frequency characteristics. Section III details the simulation experiments and EEG analysis experiments.
Section IV presents the results of these experiments, and Section V provides related discussions. Finally, Section VI offers the conclusions of this study.

\section{Methods}

\subsection{Multivariate Scale Mixture Model of Scalp EEG Signals}
\begin{figure}[!t]
\centering
\includegraphics[width=1.0\hsize]{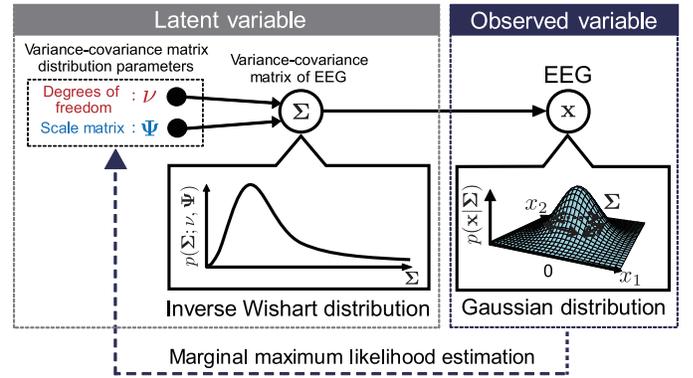}
\caption{Graphical representation of the proposed multivariate scale mixture model, which describes the stochastic relationship between EEG signals and its variance-covariance matrix.
The white nodes are random variables, and the black nodes are parameters to be estimated.
In the model, an EEG signal $\mathbf{x} \in \mathbb{R}^{D}$ is handled as a random variable that follows a multivariate Gaussian distribution with mean vector $\mathbf{0}$ and a variance-covariance matrix $\mathbf{\Sigma} \in \mathbb{R}^{D \times D}$ which is also a random variable following the inverse Wishart distribution determined by the degrees of freedom $\nu \in \mathbb{R}^+$ and the scale matrix $\mathbf{\Psi} \in \mathbb{R}^{D \times D}$. The parameter $\nu$ represents stochastic fluctuations of the EEG signals.
The variance-covariance matrix distribution parameters are estimated via marginal likelihood maximization from the recorded EEG signals.}
\label{fig:model}
\end{figure}

Fig. \ref{fig:model} shows the stochastic relationship between the recorded EEG signal $\mathbf{x} \in \mathbb{R}^{D}$ ($D$ is the number of electrodes) and its variance-covariance matrix $\mathbf{\Sigma} \in \mathbb{R}^{D \times D}$ as a graphical model.
In the proposed model, $\mathbf{x}$ is handled as a random variable that follows a multivariate Gaussian distribution with mean zero and a variance-covariance matrix  $\mathbf{\Sigma}$.
The variance-covariance matrix $\mathbf{\Sigma}$ is also a random variable for which the distribution is determined by the degrees of freedom parameter $\nu \in \mathbb{R}^+$ and the scale matrix parameter $\mathbf{\Psi} \in \mathbb{R}^{D \times D}$.
Therefore, the variance-covariance matrix $\mathbf{\Sigma}$ is interpreted as a latent variable in the model.

First, the conditional distribution of EEG signal $\mathbf{x}$ given $\mathbf{\Sigma}$ is expressed via the following multivariate Gaussian distribution with a mean vector of zero:
\begin{align}
	p(\mathbf{x}|{\mathbf{\Sigma}}) &= {\mathcal N}(\mathbf{x}|\mathbf{0}, {\mathbf{\Sigma}}) \nonumber\\
&= \frac{1}{(2\pi)^{\frac{D}{2}} |\mathbf{\Sigma}|^{\frac{1}{2}}} \mathrm{exp} \left[-\frac{1}{2}\mathbf{x}^\mathrm{T} {\mathbf{\Sigma}}^{-1} \mathbf{x}\right]. \label{eq:gauss_x} 
\end{align}
The the variance-covariance matrix is assumed to obey an inverse Wishart distribution ${\mathcal {IW}}(\mathbf{\Sigma}; \nu, \mathbf{\Psi})$, which is known as a conjugate prior for the variance-covariance matrix of a multivariate Gaussian distribution~\cite{t2006}:
\begin{align}
	p(\mathbf{\Sigma}) &= {\mathcal {IW}}(\mathbf{\Sigma};\nu,\mathbf{\Psi}) \nonumber\\
&= \frac{|\mathbf{\Psi}|^{\frac{\nu}{2}}}{2^{\frac{\nu D}{2}} \Gamma_D \left(\frac{\nu}{2}\right)} |\mathbf{\Sigma}|^{-\frac{\nu+D+1}{2}} \mathrm{exp} \left[-\frac{\mathrm{tr}(\mathbf{\Psi} \mathbf{\Sigma}^{-1})}{2}\right],\label{eq:p_sigma} 
\end{align}
where $\nu$ and $\mathbf{\Psi}$ determine the inverse Wishart distribution and are referred to as the degrees of freedom and the scale matrix, respectively.
Considering the marginal distribution of $\mathbf{x}$, $\mathbf{\Sigma}$ can be integrated out as follows:
\begin{align}
	p(\mathbf{x}) &= \int p(\mathbf{\Sigma})p(\mathbf{x}|\mathbf{\Sigma}) \mathrm{d}{\mathbf{\Sigma}} \nonumber \\
		&= \int {\mathcal {IW}}(\mathbf{\Sigma}; \nu, {\mathbf{\Psi}}) {\mathcal N}(\mathbf{x}|\mathbf{0}, \mathbf{\Sigma}) \mathrm{d}{\mathbf{\Sigma}} \label{eq:marginal_x} \\ 
		&= \int \frac{|{\mathbf{\Psi}}|^{\frac{\nu}{2}}}{2^{\frac{\nu D}{2}} \Gamma_D \left(\frac{\nu}{2}\right)} |\mathbf{\Sigma}|^{-\frac{\nu+D+1}{2}}\mathrm{exp} \left[-\frac{\mathrm{tr}(\mathbf{\Psi} \mathbf{\Sigma}^{-1})}{2}\right] \nonumber \\
&\quad \times\frac{1}{(2\pi)^{\frac{D}{2}} |\mathbf{\Sigma}|^{\frac{1}{2}}} \mathrm{exp} \left[-\frac{1}{2}\mathbf{x}^\mathrm{T} {\mathbf{\Sigma}}^{-1} \mathbf{x}\right] \mathrm{d}{\mathbf{\Sigma}} \nonumber \\
		&= \frac{|{\mathbf{\Psi}}|^{\frac{\nu}{2}}}{2^{\frac{\nu D}{2}} (2\pi)^\frac{D}{2} \Gamma_D \left(\frac{\nu}{2}\right)} \int |\mathbf{\Sigma}|^{-\frac{\nu+D+2}{2}}\nonumber\\
&\quad \times \mathrm{exp} \left[-\frac{1}{2}\mathrm{tr}\{(\mathbf{\Psi} + \mathbf{x} \mathbf{x}^\mathrm{T})\mathbf{\Sigma}^{-1}\}\right] \mathrm{d}{\mathbf{\Sigma}} \nonumber \\
\label{eq:p_x}
&= \frac{\Gamma(\frac{\nu+1}{2})}{\Gamma(\frac{\nu-D+1}{2})} \frac{|\frac{1}{\nu-D+1} {\mathbf{\Psi}}|^{-\frac{1}{2}}}{\left[\pi(\nu-D+1) \right]^{\frac{D}{2}}} (1+\Delta)^{-\frac{\nu+1}{2}} ,
\end{align}
where $\Delta$ is the square of the Mahalanobis distance, as follows:
\begin{align}
	\Delta &= \mathbf{x}^\mathrm{T}{\mathbf{\Psi}}^{-1}\mathbf{x}.
\end{align}

From (\ref{eq:gauss_x}) and (\ref{eq:marginal_x}), $p(\mathbf{x})$ is obtained by summing an infinite number of multivariate Gaussian distributions having different variance-covariance matrices, meaning it can be interpreted as the scale mixture of multivariate Gaussian distributions.
As stated above, the marginal distribution of multidimensional EEG signals can be modeled based on a multivariate scale mixture distribution.

\subsection{Parameter Estimation Based on Marginal Maximum Likelihood}
Let us consider the estimation of $\nu$ and $\mathbf{\Psi}$, given $N$ samples of EEG signals $\mathbf{X} = \{\mathbf{x}_n \in \mathbb{R}^{D}; n=1,2,\cdots,N \}$. The model parameters can be estimated by maximizing the marginal likelihood $p(\mathbf{X}) = \prod_{n=1}^{N} p(\mathbf{x}_n)$.
However, obtaining the maximum likelihood solution of a marginal likelihood is generally complex, and thus it is difficult to optimize the solution analytically~\cite{t2006}.
Therefore, we conduct this optimization for $\nu$ and $\mathbf{\Psi}$ based on the expectation-maximization (EM) algorithm~\cite{Models1998}.
The EM algorithm is iterated via application of an expectation step (E-step) and a maximization step (M-step).

To simplify (\ref{eq:p_x}), we define the new parameters as:
\begin{align}
	\label{eq:eq6}
	\nu&= \nu' + D - 1, \\
	\label{eq:eq7}
	\mathbf{\Psi}&= \nu' \mathbf{\Psi}'.
\end{align}
Accordingly, the marginal distribution can be expressed as
\begin{align}
\label{eq:eq8}
p(\mathbf{x}_n) &=\int \mathcal{IW}(\mathbf{\Sigma}_n; \nu'+D-1, \nu' \mathbf{\Psi}') \mathcal{N}(\mathbf{x}_n|\mathbf{0}, \mathbf{\Sigma}_n) \mathrm{d} {\mathbf{\Sigma}_n} \\
\label{eq:eq9}
&=\frac{\Gamma(\frac{\nu'+D}{2})}{\Gamma(\frac{\nu'}{2})} \frac{|{\mathbf{\Psi}'}|^{-\frac{1}{2}}}{\left(\pi \nu' \right)^{\frac{D}{2}}} \left(1+\frac{\Delta '}{\nu '} \right)^{-\frac{\nu'+D}{2}},
\end{align}
where
\begin{align}
	\Delta ' = \mathbf{x}_n^\mathrm{T} ({\mathbf{\Psi} '})^{-1} \mathbf{x}_n.
\end{align}
Equation (\ref{eq:eq9}) is equivalent to multivariate Student-$t$ distribution $\mathrm{St}(\mathbf{x}_n|\nu', \mathbf{\Psi}')$~\cite{t2006}.
Then, we redefine the latent variable and replace (\ref{eq:eq8}) with the following equivalent expression, thereby allowing an efficient calculation (refer to Appendix).

\begin{equation}
\label{eq:eq11}
		p(\mathbf{x}_n) = \int \mathrm{IG}(\tau_n;\nu'/2,\nu'/2) \mathcal{N}(\mathbf{x}_n|\mathbf{0}, \tau_n \mathbf{\Psi}') \mathrm{d}{\tau_n},
\end{equation}
where $\tau_n$ is a new latent variable following an inverse Gamma distribution $\mathrm{IG(\cdot)}$.
The model parameters $\nu'$ and $\mathbf{\Psi}'$ are estimated by maximizing the marginal likelihood as outlined below.

\begin{enumerate}
\setlength{\parskip}{0cm}
\setlength{\itemsep}{0cm}
\item[(i)] Initialize each parameter by selecting arbitrary starting values.
\item[(ii)] \textit{E-step}. Calculate the expectation of the complete-data log-likelihood, denoted as $Q(\nu',\mathbf{\Psi}')$.
\begin{align}
  Q(\nu',&\mathbf{\Psi}') \notag \\
  &=\mathbb{E} \left[\ln{\prod_{n=1}^{N}\mathrm{IG}(\tau_n;\nu'/2,\nu'/2) \mathcal{N}(\mathbf{x}_n|\mathbf{0}, \tau_n \mathbf{\Psi}')}\right]  \notag  \\
  &=\sum_{n=1}^{N} \biggl[-\frac{D}{2}\ln{(2\pi)}-\frac{D}{2}\mathbb{E}\left[\ln{\tau_n}\right]-\frac{1}{2}\ln{|\mathbf{\Psi}'|} \notag \\
  &\quad-\frac{1}{2}\mathbb{E}\left[\tau^{-1}_n\right]\Delta'+\frac{\nu'}{2} \ln{\frac{\nu'}{2}} - \ln{\Gamma{\left(\frac{\nu'}{2}\right)}} \notag \\
  &\quad- \left(\frac{\nu'}{2}+1\right)\mathbb{E}\left[\ln{\tau_n}\right]-\frac{\nu'}{2}\mathbb{E}\left[\tau^{-1}_n \right] \biggr],
\end{align}
where $\mathbb{E}\left[\tau^{-1}_n\right]$ and $\mathbb{E}\left[\ln{\tau_n}\right]$ are derived by calculating the posterior distribution $p(\tau_n|\mathbf{x}_n)$ of the latent variable $\tau_n$ as follows:
\begin{equation}
\mathbb{E}\left[\tau^{-1}_n\right]=\frac{\nu'+D}{\nu'+\Delta'},
\end{equation}
\begin{equation}
\mathbb{E}\left[\ln{\tau_n}\right]=-\ln{\mathbb{E}\left[\tau^{-1}_n\right]}+\ln{\left(\frac{\nu'+D}{2}\right)}-{\psi}\left(\frac{\nu'+D}{2}\right),
\end{equation}
where $\psi(\cdot)$ is a digamma function.
\item[(iii)] \textit{M-step}. Update the parameters by maximizing $Q(\nu',\mathbf{\Psi}')$.
By setting the derivative of $Q(\nu',\mathbf{\Psi}')$ with $\mathbf{\Psi}'$ equal to zero, the new scale matrix is obtained as

\begin{equation}
{^\mathrm{new} \mathbf{\Psi}}' = \frac{1}{N} \sum_{n=1}^{N} \mathbb{E}\left[\tau^{-1}_n\right] \mathbf{x}_n \mathbf{x}_n^\mathrm{T}.
\end{equation}
Because there is no closed-form expression for the degrees of freedom parameter $\nu'$, we estimate $\nu'$ by iteratively maximizing $Q(\nu',\mathbf{\Psi}')$ using the bisection method.

\begin{equation}
{^\mathrm{new} \nu}' = \argmax_{\nu'}  Q(\nu', {^\mathrm{new} \mathbf{\Psi}}').
\end{equation}
\item[(iv)] Evaluate the log-likelihood $\ln{p(\mathbf{X})}$ and repeat steps (ii)--(iv) until the calculation converges. Finally, estimated parameters $\nu'$ and $\mathbf{\Psi}'$ are transformed to variance-covariance matrix distribution parameters $\nu$ and $\mathbf{\Psi}$.
\end{enumerate}
Using these procedures, the parameters of the proposed model can be estimated from the recorded EEG signals.
Here, the parameter $\nu$ corresponds to the degrees of freedom of the multivariate Student-$t$ distribution parameter (refer to (\ref{eq:eq6})); hence, $\nu$ controls the Gaussianity of the distribution.
In the framework of the scale mixture model, the Gaussianity of EEG is considered to change owing to the stochastic fluctuation of the scale parameter (i.e., the variance-covariance matrix).
Therefore, the fluctuation of the variance-covariance matrix can be evaluated by estimating $\nu$ from recorded EEG signals.
 
\subsection{Proposed EEG Analysis Methods}
\begin{figure*}[!ht]
\centering
\includegraphics[width=0.75\hsize]{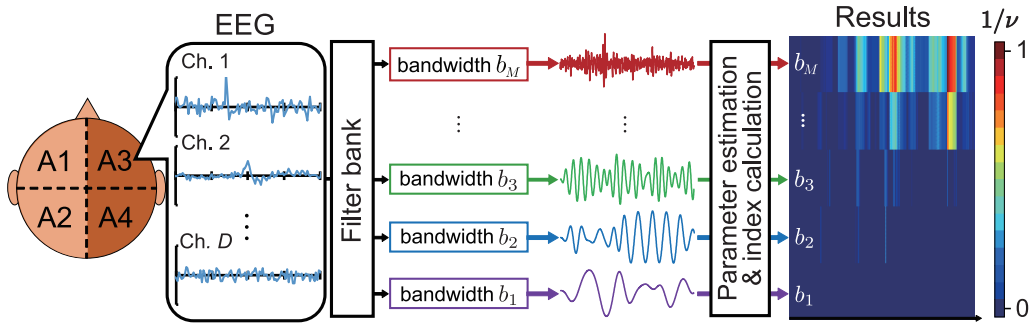}
\caption{Overview of the proposed analysis method. The recorded EEG signals are decomposed into multiple frequency bands ($b_1$--$b_M$) using a filter bank consisting of parallel band-pass filters. 
The evaluation feature $1/\nu$ is then calculated for each frequency band, based on the parameter estimation results of the proposed model.
The results are shown as a color map.}
\label{fig:system}
\end{figure*}
Fig.~\ref{fig:system} shows the overall outline of the proposed analysis method.
In the proposed method, observed EEG signals are decomposed into multiple frequency bands by using a filter bank consisting of parallel band-pass filters.
Additionally, we estimate non-Gaussianity for signals in each frequency band based on the proposed multivariate scale mixture model.

First, the EEG signal at time $t$ recorded from the $D$ pair of electrodes is defined as $\mathbf{x}_t \in \mathbb{R}^D$. 
Then, $\mathbf{x}_t$ is divided into $M$ frequency bands ($b_1, \cdots, b_M$) by applying a filter bank consisting of a third-order Butterworth band-pass filter; the obtained signal is defined as $x_t^{(b_m)}$ ($m = 1,\cdots, M$).
Second, parameter estimation for the variance-covariance matrix distribution based on the proposed model is performed on $\mathbf{x}^{(b_m)}_t$ for each frequency band, and $\nu$ characterizing the stochastic fluctuation is obtained.
Because the characteristics of EEG signals change significantly in a time series, a sliding window of length $W$ (s) is applied to $\mathbf{x}^{(b_m)}_t$ in each band, and $\nu$ is estimated from the sample in the window following the procedure described in Section II-\textit{B}.
The time window is slid by $S$ (s) continuously, resulting in the estimation of $ \nu$ in a time series.

Here, $\nu$ is a parameter that controls the Gaussianity of EEG signals.
The EEG distribution becomes closer to a Gaussian form as the value of $\nu$ approaches $\infty$; that is, the stochastic fluctuations decrease as $\nu$ increases.
In this paper, we calculate $1/\nu$, which is the reciprocal of $\nu$, to provide intuitive meaning and to serve as a feature that characterizes the non-Gaussianity. 
The feature $1/\nu$ indicates that the larger its value, the larger the stochastic fluctuation of the EEG signal.

From the above, the stochastic fluctuations latent in each frequency band of EEG signals can be estimated continuously.
It is also possible to obtain the spatial distribution of stochastic fluctuations by dividing the EEG electrode arrangement into multiple regions in advance and performing the above analysis for each region.

\section{Experiments}
\begin{figure}[!t]
\centering
\includegraphics[width=0.5\hsize]{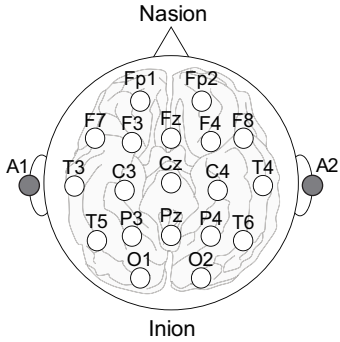}
\caption{International 10--20 electrode montage}
\label{fig:electrodes}
\end{figure}

\subsection{Simulation}
To verify the accuracy of parameter estimation based on the proposed model, we performed a simulation experiment to evaluate the error rate between the true and estimated values of the parameters.
Since the marginal distribution of the proposed model is equivalent to the multivariate Student-$t$ distribution, a random number sequence $\{\mathbf{x}_t \in \mathbb{R}^D; t = 1, \cdots, T\}$ following a Student-$t$ distribution $\mathrm{St}(\mathbf{x}_t|\nu'_0, \mathbf{\Psi}'_0)$ was generated.
Here, the $\{\mathbf{x}_t \}$ values were regarded as a time series of an EEG signal recorded at a sampling frequency of $f_s$.
The accuracy of the distribution estimation was verified by comparing the true values $\nu_0$ and $\mathbf{\Psi}_0$ with estimated values $\nu$ and $\mathbf{\Psi}$, after converting $\nu'_0$ and $\mathbf{\Psi}'_0$ to inverse Wishart distribution parameters based on (\ref{eq:eq6}) and (\ref{eq:eq7}).
As an index of estimation accuracy for each parameter, the absolute percentage error was defined as ${|\nu_0 -\nu|}/{|\nu_0|}\times100$ and ${\|\mathbf{\Psi}_0-\mathbf{\Psi}\|_F}/{\|\mathbf{\Psi}_0\|_F}\times100$.
$\|\mathbf{\Psi}\|_F$ is the Frobenius norm~\cite{GeneHowardGolub2013} of $\mathbf{\Psi}$ and is obtained as
\begin{align}
	\|\mathbf{\Psi}\|_F=\sqrt{\sum_{i=1}^D \sum_{j=1}^D |{\psi_{ij}}|^2},
\end{align}
where ${\psi_{ij}}$ is the element of $\mathbf{\Psi}$.
In the estimation of each parameter, the first $W$ values of the signals $\{\mathbf{x}_t \}$ were used.
The window length $W$ took values of 1, 2, 5, 10, 15, 20, 30, 50, and 100 s, and the number of dimensions $D$ took values of 1, 2, 4, 8, 16, and 19.
The average absolute percentage errors were calculated by changing the true values of multivariate $t$-distribution parameters 400 times ($\nu'_0 = 0.5, 1.0, 1.5, \cdots, 10.0, {\psi'_0}_{ii} = 1.0, 2.0, 3.0, \cdots, 20.0$).
Note that $\mathbf{\Psi}'_0$ was changed only in terms of its diagonal components ${\psi'_0}_{ii}$, and the off-diagonal components were fixed at 0.5.
To evaluate the calculation cost of the proposed analysis method, the computation time required for each parameter estimation was measured simultaneously.
The $T$ and $f_s$ values in artificial data generation were set as 100 s and 500 Hz.
The computer used in the experiment was an Intel Core (TM) i7–6900K (3.2 GHz), 64.0 GB RAM.

\subsection{EEG analysis}
Experimental analyses were conducted to evaluate the validity of the proposed model and the effectiveness of the proposed feature $1/\nu$ for epileptic seizure detection.
Twenty epileptic patients with focal epilepsy participated in the experiments.
Table~\ref{table:conditions} summarizes patient information, analysis times, and duration of the epileptic seizures.
The EEG signals were recorded with a digital sampling frequency of 500 Hz using an electroencephalograph system (Neurofax EEG-1218, Nihon Kohden, Tokyo, Japan) while the patients were in the supine position.
The 19-channel surface electrodes ($D=19$) were placed on the scalp according to the international 10--20 electrode system, with reference electrodes on both earlobes: A1 and A2 (see Fig. \ref{fig:electrodes}).
The experiments were approved by the Okayama University Ethics Committee (approval No: 1706-019). The onset and offset of a focal seizure in each EEG recording were marked by a board-certified epileptologist (T.A.).

\begin{table}[!t]
\centering
 \caption{Patient conditions}
 \label{table:conditions}
  \begin{tabular}{lllll}
   \toprule 
   \multirow{2}{*}{Patient} &  \multirow{2}{*}{Sex} &  Age  & Total data & Seizure \\
   & & (year) &  length (s) & duration (s) \\
   \midrule 
   A & Male & 2 & 300 & 71\\ 
   B & Male & 23 & 300 & 54\\ 
   C & Female & 4 & 300 & 31\\ 
   D & Male & 4 & 380 & 93\\ 
   E & Male & 0.5 & 320  & 39\\ 
   F & Male & 41 & 300 & 32\\
   G & Female & 3 & 240 & 53\\
   H & Male & 19 & 390 & 36\\ 
   I & Male & 0.8 & 390  & 98\\ 
   J & Male & 20 & 360 & 16\\ 
   K & Male & 36 & 300 & 23\\ 
   L & Male & 9 & 300 & 43\\ 
   M & Male & 13 & 300  & 43\\ 
   N & Male & 15 & 300 & 48\\
   O & Male & 8 & 300 & 17\\ 
   P & Female & 19 & 300 & 69\\ 
   Q & Male & 27 & 420 & 62\\ 
	 R & Male & 38 & 300 & 65\\ 
	 S & Male & 17 & 300 & 17\\ 
	 T & Male & 19 & 300 & 65\\ 
   \bottomrule 
  \end{tabular}
\end{table}

First, the proposed model was fitted to the recorded EEG signals for all participants, and the feature $1/\nu$ was calculated.
We set the frequency bands in the proposed method to $b_m \in\{\delta, \theta, \alpha,\beta, \gamma\}$, and decomposed EEG signals into $\delta$ (1--3 Hz), $\theta$ (4--7 Hz), $\alpha$ (8--12 Hz), $\beta$ (13--24 Hz), and $\gamma$ (25--100 Hz).
These frequency bands are generally used to extract features of EEG signals~\cite{ep1994}.
Here, the fitting of the model and the calculation of $1/\nu$ were performed continuously using a sliding window with length $W$ = 15 s and sliding width $S$ = 1 s.

We then performed a goodness-of-fit test to validate the proposed model for EEG signals in each frequency band.
As an evaluation index for the goodness-of-fit, we used the Bayesian information criterion (BIC)~\cite{Schwarz1978}, which balances the fitness and complexity of the model.
\begin{align}
	\mathrm{BIC} = -2 \mathrm{ln}L(\hat{\theta}) + k \mathrm{ln}(N_W).
\end{align}
Here, $\mathrm{ln}L(\hat{\theta})$ is the log-likelihood of the model, $k$ is the number of estimated parameters in the model, and $N_W$ is the sample size in the sliding window. 
BIC was calculated for the fitting results for each sliding window.
For comparison, BIC was also calculated for a multivariate Gaussian and Cauchy distribution models, having heavier tail than the Gaussian distribution.

Next, to verify the effectiveness of the evaluation of EEG non-Gaussianity based on the proposed method, we investigated whether epileptic seizures and non-seizures were classified accurately by using the calculated feature $1/\nu$ of each frequency band ($\delta$--$\gamma$ bands). 
The calculated $1/\nu$ results for each patient were divided into those obtained from seizure and non-seizure segments.
Here, a non-seizure segment is defined as the period from the start of the recording to the onset of seizures.
Non-seizure segments that include extraneous noise owing to unexpected body movements or electrode shifts in the original EEG signals were excluded from the analysis.
In addition, because the sample size for calculating $1/\nu$ differed significantly between the seizure and non-seizure segments, the calculated $1/\nu$ in the non-seizure segment was randomly sampled based on the sample size for calculating $1/\nu$ in the seizure segment, to obtain uniform sample size for each patient.

As an evaluation index of classification performance, we calculated the area under the receiver operating characteristic curve (AUC) based on the receiver operating characteristic (ROC) analysis.
The AUC is an evaluation scale calculated from the ROC curve plotting the relationship between the false positive and true positive rates.
The closer the AUC value is to 1, the higher the classification performance.

For comparison, the AUC was calculated in the same manner using the three time-domain features conventionally used for EEG-based epileptic seizure detection: root mean square (RMS), absolute value of third-order cumulant ($|$ToC$|$), and approximate entropy (ApEn).
\begin{itemize}
  \item RMS is a common feature that characterizes the amplitude information of EEGs~\cite{Hamedi2014} and can be calculated as follows:
    \begin{equation} 
    		\mathrm{RMS} = \sqrt{\frac{1}{N_W} \sum_{i=1}^{N_W} (x_i)^2},
    \end{equation}
    where $x_i$ is the EEG signal at an arbitrary electrode.
  \item ToC is a higher-order statistic and is used to investigate nonlinear variations in the signal.
    For zero-mean signals, the ToC is the same as the third-order moments (i.e., skewness), which is a measure of the asymmetry of the probability distribution~\cite{Chua2010-qq}.
    The effectiveness of ToC in epileptic seizure detection has been reported by earlier works~\cite{Sharma2020-ko, Sharma2020-rg}.
    In this experiment, the absolute value of ToC, $|$ToC$|$, was calculated.
  \item ApEn proposed by Pincus~\cite{Pincus1991} measures the regularity and unpredictability of fluctuations over a time-series signal.
    Smaller values of ApEn indicate strong regularity in a data sequence.
    The ApEn is also a commonly used feature for epileptic seizure detection, and ApEn of EEG is shown to decrease with epileptic seizures~\cite{Srinivasan2007-yw,Ocak2009-rh,Arunkumar2018-cq}.
    The embedded dimension $m$ and vector comparison distance $r$ in ApEn were set to recommended values~\cite{Pincus1991,Richman2000}; $m = 2$ and $r = 0.2sd$, where $sd$ is the standard deviation of the data.
\end{itemize}
These conventional features were calculated continuously using a sliding window with the same settings as those used in calculating $1/\nu$.
The frequency decomposition before the calculation of each feature was also conducted using the same filter bank ($\delta$--$\gamma$) as the proposed method.
In addition, although the proposed feature is defined for an arbitrary set of electrodes, the conventional features are calculated for each electrode.
Therefore, the conventional features were calculated for the Cz channel (see Fig.~\ref{fig:electrodes}) obtained from the top of the head.
Other experimental conditions were the same as the simulation experiment.

\begin{figure}[t]
\centering
\includegraphics[width=1.0\hsize]{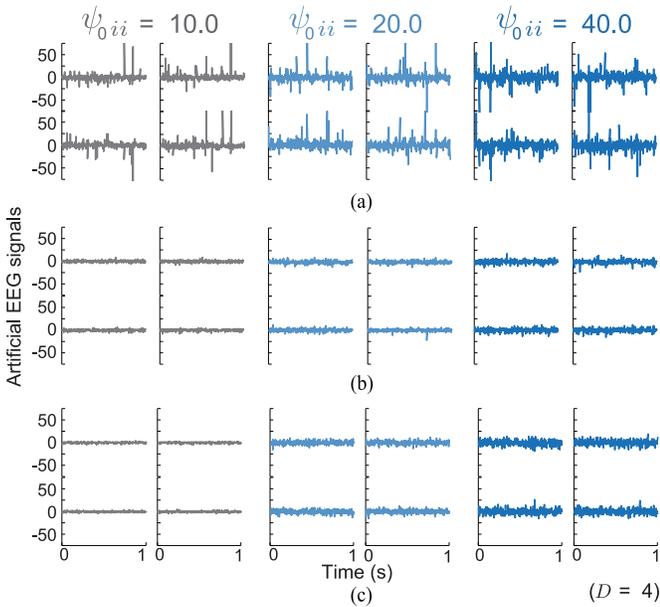}
\caption{Examples of artificially generated EEG signals ($D$ = 4) with $\nu_0$ set to (a) $\nu_0 = 5.0$, (b) $\nu_0 = 8.0$, and (c) $\nu_0 = 13.0$; in each case, ${{\psi_0}_{ii}}$ was set to ${{\psi_0}_{ii}} = 10.0$, ${{\psi_0}_{ii}} = 20.0$, and ${{\psi_0}_{ii}} = 40.0$.}
\label{fig:sim_EEG}
\end{figure}

\begin{figure}[t]
\centering
\includegraphics[width=0.75\hsize]{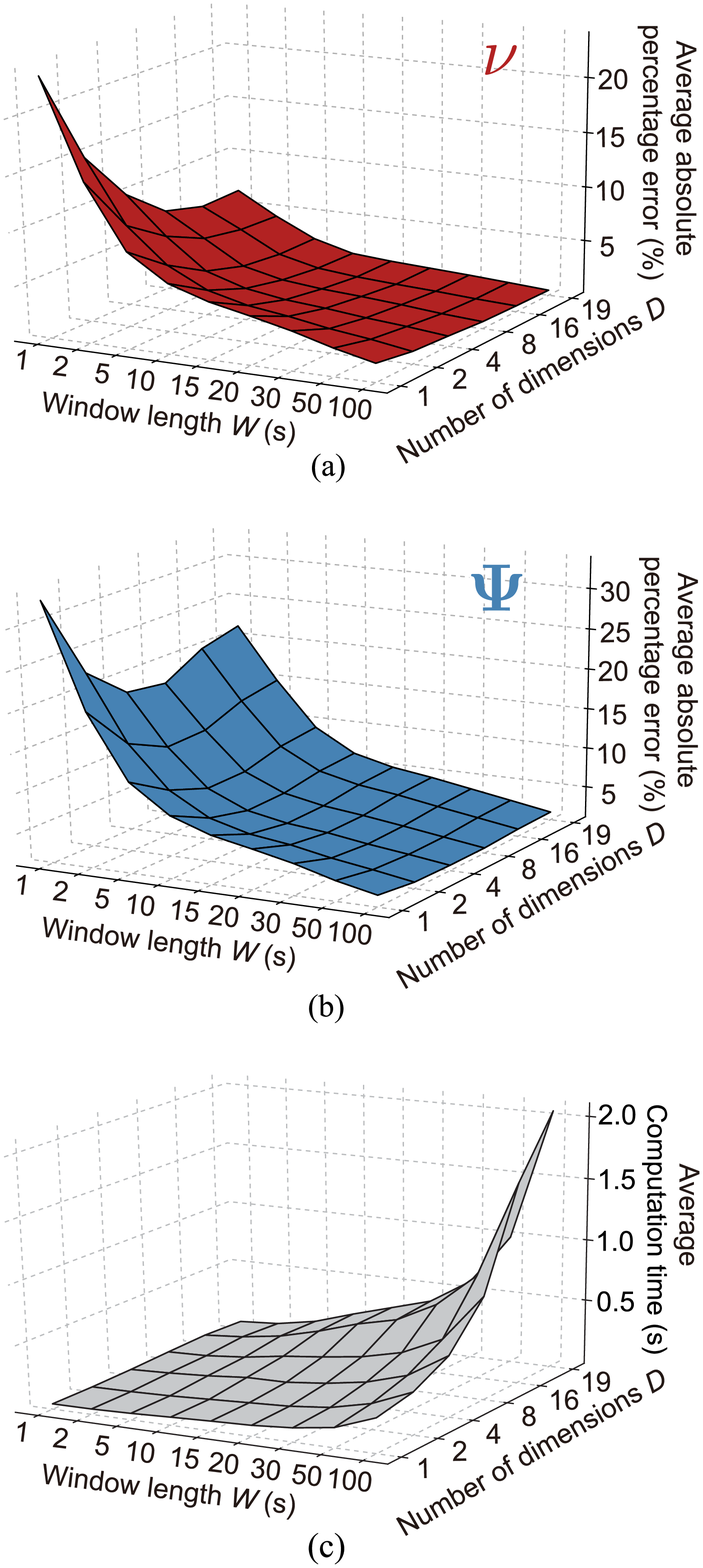}
\caption{Average absolute percentage errors and computation time for each combination of the number of input dimensions ($D$) and sliding window length ($W$) in the estimation of variance-covariance distribution parameters. (a) Average absolute percentage error for degrees of freedom $\nu$. (b) Average absolute percentage error for scale matrix $\mathbf{\Psi}$. (c) Average computation time.}
\label{fig:ES_param}
\end{figure}

\section{Results}

In the simulation experiments, we generated time-series simulation data and verified the estimation accuracy of the distribution parameter.
Fig. \ref{fig:sim_EEG} shows examples of time-series waveforms of $\mathbf{x}_t$, where the parameters of the inverse Wishart distribution used in artificial data generation were set as $\nu_0$ = 5.0, $\nu_0$ = 8.0, and $\nu_0$ = 13.0, and ${\psi_0}$ was changed to ${\psi_0}_{ii}$ = 10.0, ${\psi_0}_{ii}$ = 20.0, and ${\psi_0}_{ii}$ = 40.0 for each $\nu_0$.
In the examples, the number of dimensions for artificial EEG signals is set to $D = 4$, and the signals for each dimension are shown.
The vertical and horizontal axes indicate the signal values and time, respectively. 
Fig. \ref{fig:ES_param}(a) and (b) show the average absolute percentage errors in the estimation of $\nu$ and $\mathbf{\Psi}$ resulting from changing the window length $W$ and the number of dimensions $D$.
The average computation time for each estimation is shown in Fig.~\ref{fig:ES_param}(c).

\begin{figure*}[ht]　
\centering
\includegraphics[width=0.9\hsize]{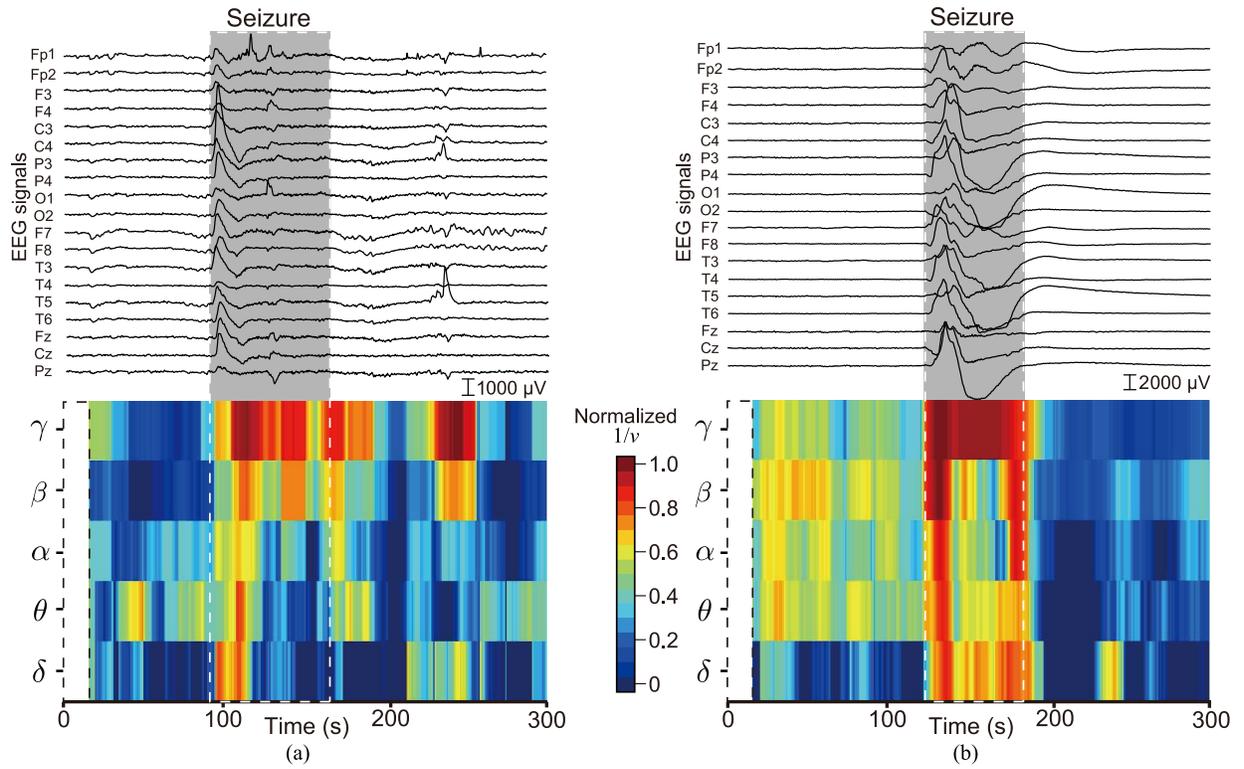}
\caption{Raw EEG signals and corresponding analysis results from the proposed method for (a) patient A and (b) patient B.
Proposed feature $1/\nu$ was normalized based on min-max normalization, rescaling the range of features to [0, 1].}
\label{fig:Colormap}
\end{figure*}

In the EEG analysis experiments, we performed a goodness-of-fit test for the EEG signals and evaluated non-Gaussianity using the proposed feature.
Table~\ref{table:BIC} lists the percentage of times that the BIC of each model became a minimum in the EEG signals of the $\delta$--$\gamma$ bands.
%
\begin{table}[!t]
\centering
 \caption{Percentage of times each model was selected for different frequency bands based on BIC}
  \label{table:BIC}
 \begin{threeparttable}
  \begin{tabular}{llll}
   \toprule 
    & \multicolumn{3}{c}{Model}\\ \cmidrule(r){2-4}
   {Frequency band} & {Proposed} & \; Gaussian & Cauchy \\
   \midrule 
   $\gamma$ (25--100 Hz) & 99.28\%  &\; 0.21\% *** &0.51\% ***\\ 
   $\beta$ (13--24 Hz)  & 99.05\%  &\; 0.59\% *** &0.36\% ***\\ 
   $\alpha$ (8--12 Hz) & 96.79\%  &\; 3.09\% *** &0.12\% ***\\ 
   $\theta$ (4--7 Hz) & 95.47\%  &\; 4.42\% *** &0.11\% ***\\ 
   $\delta$ (1--3 Hz) & 81.80\%  &\; 17.92\% *** &0.28\% ***\\ 
   \bottomrule 
   \addlinespace[1.0mm]
  \end{tabular}
    \begin{tablenotes}[para,flushleft]
  ***: significant difference with the proposed model as indicated by a McNemar test with a Holm adjustment (\textit{p} $<$ 0.001)
  \end{tablenotes}
  \end{threeparttable}
\end{table}
%
The table also lists the McNemar test results (significance level: 0.1\%) adjusted by the Holm method using the proposed model as a control group.
In all frequency bands, significant differences were observed between the proposed and other models ($p < 0.001$).
In Fig. \ref{fig:Colormap}, color maps show the $1/\nu$ calculations corresponding to the raw EEG waveforms of patients A and B.
%

In the color maps, $1/\nu$ was normalized to ensure that the maximum and minimum values of all frequency bands would be 1, and 0, respectively.
The shaded areas in the waveforms and the areas surrounded by white dotted lines indicate epileptic seizure occurrences diagnosed by an epileptologist.

\begin{figure}[!t]　
\centering
\includegraphics[width=1.00\hsize]{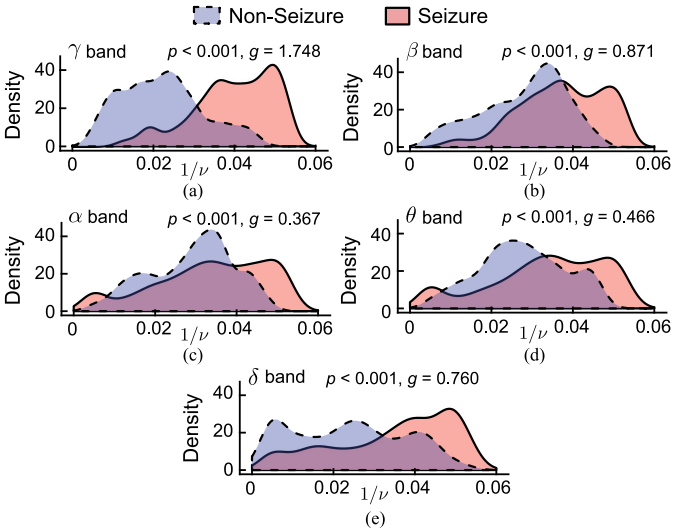}
\caption{Experimental $1/\nu$ distributions estimated from EEG signals. The \textit{p}-value from the paired $t$-test and effect size \textit{g} are also shown. (a) $\gamma$ band. (b) $\beta$ band. (c) $\alpha$ band. (d) $\theta$ band. (e) $\delta$ band.}
\label{fig:dens}
\end{figure}

Fig. \ref{fig:dens} shows the $1/\nu$ distribution in each frequency band for all patients, as calculated by kernel density estimation~\cite{Parzen1962} for each seizure and non-seizure segment.
The figure also shows the results of the paired $t$-test (significance level: 0.1\%) and the effect size $g$~\cite{Hedges1981} between the $1/\nu$ distributions from seizure and non-seizure segments.
The effect size $g$ is a statistical index indicating the magnitude of the difference between the mean values of the two distributions.
In general, $0.2 \leq g < 0.5$ is interpreted as a small effect size, $0.5 \leq g < 0.8$ indicates medium effect size, and $0.8 \leq g$ indicates large effect size~\cite{Cohen2013}.

\begin{figure}[t] 
\centering
\includegraphics[width=0.75\hsize]{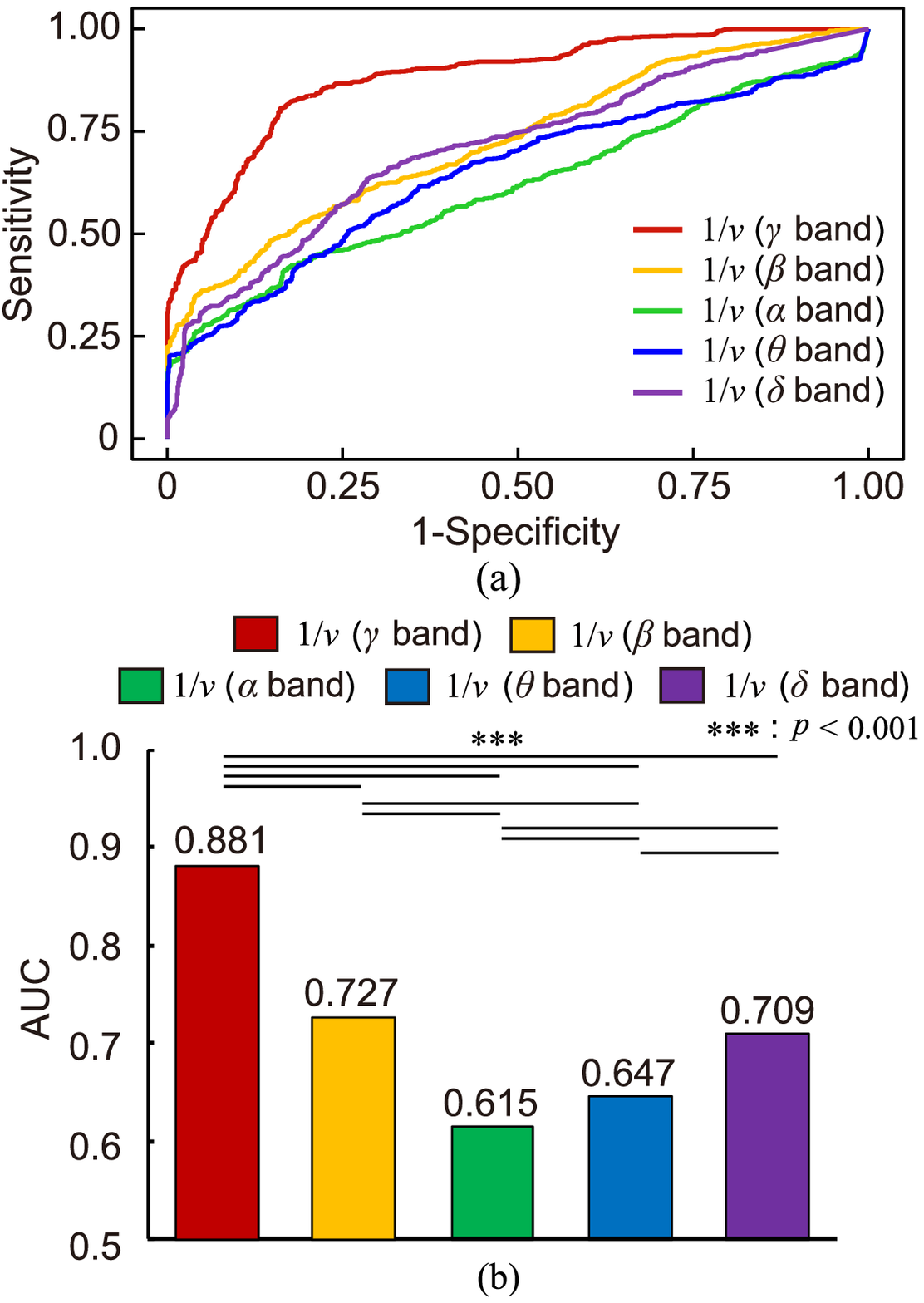}
\caption{Results of ROC analysis based on $1/\nu$ for each frequency band.
(a) ROC curves for seizure and non-seizure segments.
(b) Corresponding AUCs.
Statistical test results using DeLong test for two correlated ROC curves with the Holm adjustment are also shown ($p<0.001$).}
\label{fig:roc}
\end{figure}

\begin{figure}[t] 
\centering
\includegraphics[width=0.75\hsize]{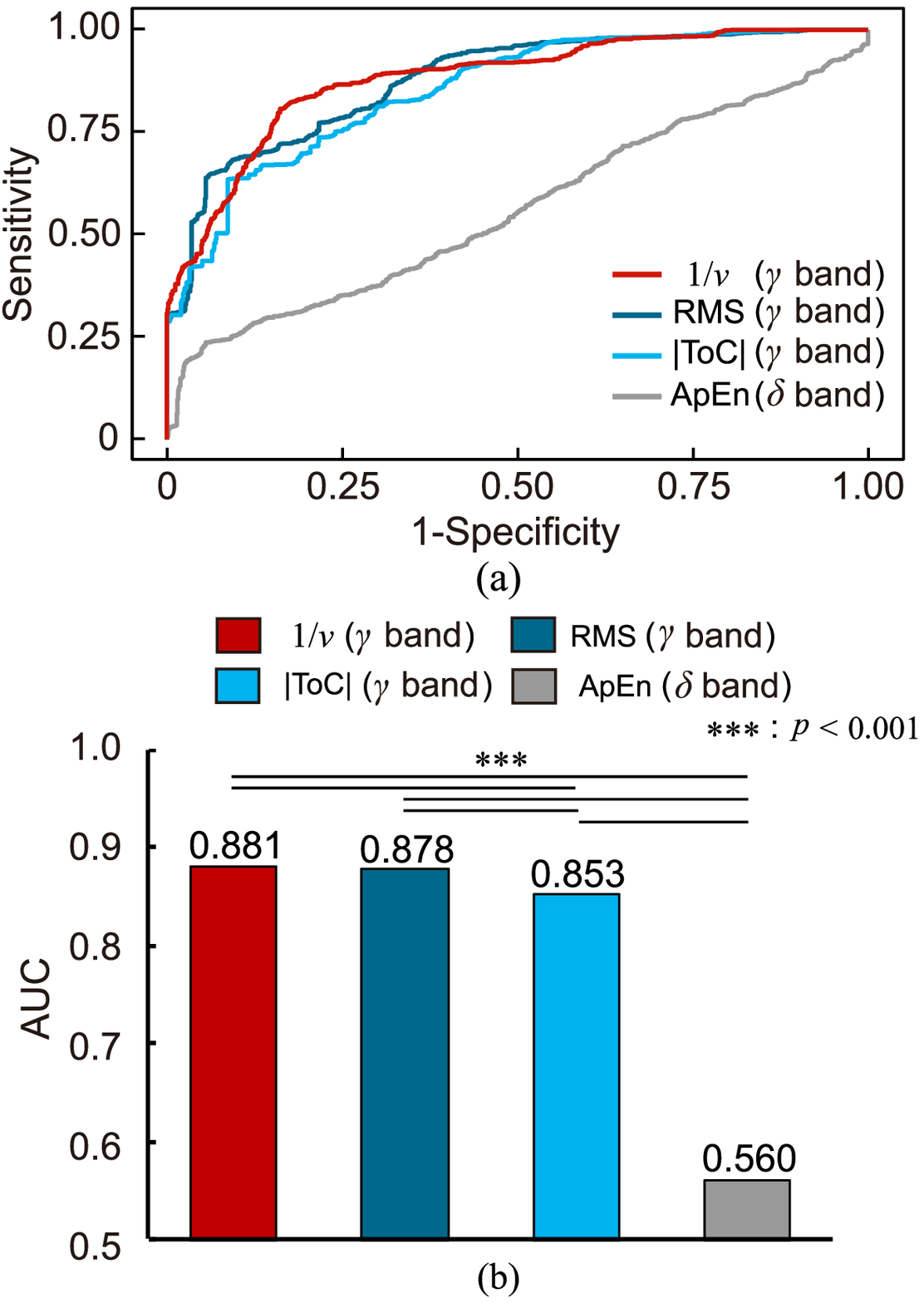}
\caption{Results of ROC analysis based on each feature for the frequency band having the best AUC.
(a) ROC curves for seizure and non-seizure segments.
(b) Corresponding AUCs.
Statistical test results using DeLong test for two correlated ROC curves with the Holm adjustment are also shown ($p<0.001$).}
\label{fig:roc_best}
\end{figure}

\begin{figure}[!t] 
\centering
\includegraphics[width=0.88\hsize]{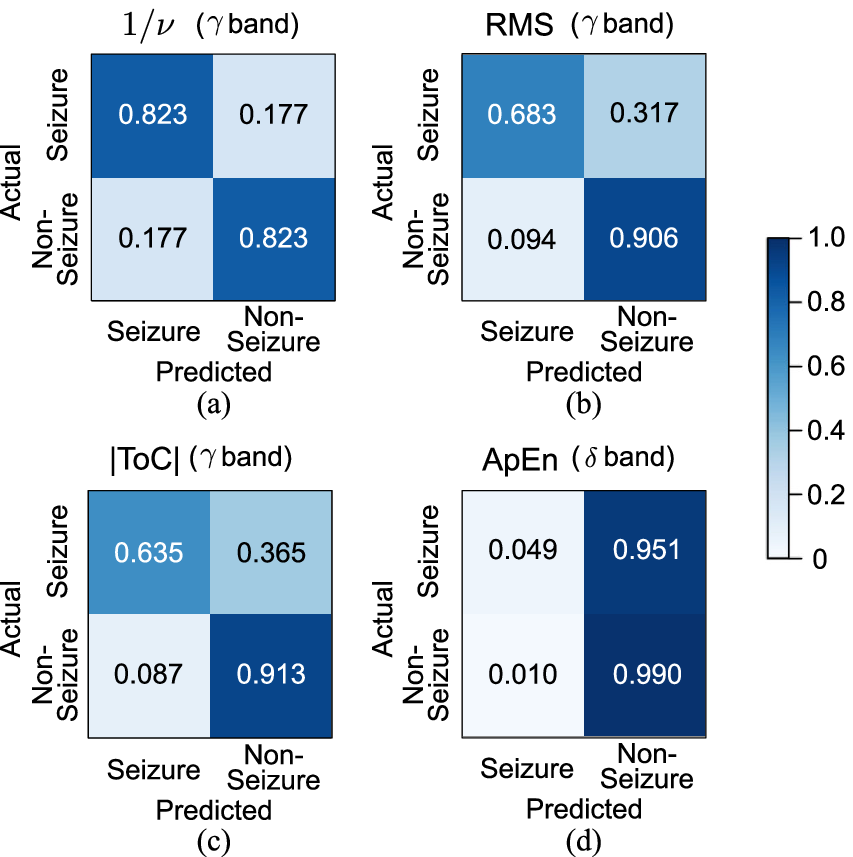}
\caption{Normalized confusion matrix given by ROC analysis. Each feature represents the band showing the best AUC. (a) $1/\nu$ in the $\gamma$ band. (b) RMS in the $\gamma$ band. (c) $|$ToC$|$ in the $\gamma$ band. (d) ApEn in the $\delta$ band. The diagonal number represents the normalized number of cases where the predicted label is equal to the actual label.}
\label{fig:conf_mat}
\end{figure}

%
\begin{table}[!t]
\centering
  \caption{AUC of all features for different frequency bands}
  \label{table:AUC}
  \begin{threeparttable}
  \begin{tabular}{lllll}
    \toprule 
    & \multicolumn{4}{c}{Features}\\ \cmidrule(r){2-5}
    {Frequency band} & {$1/\nu$} &  RMS &$|$ToC$|$ & ApEn \\
    \midrule 
    $\gamma$ (25--100 Hz) & \textbf{0.881}  & 0.878 &0.853 &0.552\\ 
    $\beta$ (13--24 Hz)  & 0.727  & 0.822 &0.786 &0.548\\ 
    $\alpha$ (8--12 Hz) & 0.615  & 0.817 &0.773 &0.555\\ 
    $\theta$ (4--7 Hz) & 0.647  & 0.746 &0.717 &0.557\\ 
    $\delta$ (1--3 Hz) & 0.709  & 0.721 &0.757 &0.560\\ 
    \bottomrule 
    \addlinespace[1.0mm]
  \end{tabular}
  \end{threeparttable}
\end{table}
%
Table~\ref{table:AUC} summarizes the AUCs of each feature based on the results of the ROC analysis for different frequency bands.
The highest AUC for each feature was $1/\nu$ in the $\gamma$ band (0.881), RMS in the $\gamma$ band (0.878), $|$ToC$|$ in the $\gamma$ band (0.853), and ApEn in the $\delta$ band (0.560).
The pairwise comparison for all AUCs in the table was performed using DeLong test with the Holm adjustment (significance level: 0.1\%).
The ROC curves obtained via seizure and non-seizure classification and the statistical test results for the AUCs are partially presented in Fig.~\ref{fig:roc} and \ref{fig:roc_best}.
Fig.~\ref{fig:roc} shows the results for each frequency band of $1/\nu$. 
The AUC of $1/\nu$ in the $\gamma$ band was significantly higher than that of $1/\nu$ in the other bands.
Fig.~\ref{fig:roc_best} shows the results of each feature for the frequency band having the best AUC.
The AUC of $1/\nu$ in the $\gamma$ band was significantly higher than that of $|$ToC$|$ and ApEn in the $\gamma$ and $\delta$ bands, respectively.
%
%
The confusion matrix for each feature obtained from the ROC analysis is shown in Fig.~\ref{fig:conf_mat}.
In the figure, each feature is for the band showing the best AUC.
The rows and columns correspond to the epileptologist assessments (actual label) and the classification results based on ROC analysis (predicted label), respectively.
The first column represents the true positives and false positives, and the second column represents the false negatives and true negatives.


\section{Discussion}
In the simulation experiment, as the value of $\nu_0$ changes from $\nu_0 = 5.0$ to $\nu_0 = 13.0$, the outlier-like variance of the waveform does not appear frequently, and the waveform stabilizes (Fig.~\ref{fig:sim_EEG}). 
This indicates that the distribution approaches a Gaussian distribution as $\nu$ increases because this parameter determines the Gaussianity of the distribution.
It can also be seen that the amplitude of the artificial data increases as the value of ${\psi_0}_{ii}$ changes from ${\psi_0}_{ii}=10.0$ to ${\psi_0}_{ii}=40.0$.
This is because $\psi_{ii}$, the diagonal component of $\mathbf{\Psi}$, is a parameter that characterizes the scale of the variance of EEG signals in each dimension.

The average absolute percentage errors in the estimation of $\nu$ and $\mathbf{\Psi}$ are approximately 2\% at $W$ = 100 s, which indicates that the estimation is accurate (Fig.~\ref{fig:ES_param}).
However, the error rate increases as the window length $W$ decreases; the error rate is approximately 25\% in $\nu$ and approximately 35\% in $\mathbf{\Psi}$ at $W$ = 1 s.
These results indicate that the estimation accuracy depends on the window length used for parameter estimation. 
When the number of dimensions was increased, the average absolute percentage error of $\nu$ decreased, and that of $\mathbf{\Psi}$ decreased initially and then increased again. 
Here, $\nu$ is a one-dimensional parameter defined for all input dimensions; therefore, a substantial increase in the number of dimensions is considered to increase the sample size used in the estimation of $\nu$, thereby decreasing the average absolute percentage error in the estimation.
The scale matrix $\mathbf{\Psi}$ is affected by the estimation result of $\nu'$ because $\mathbf{\Psi} = \nu' \mathbf{\Psi}'$; therefore, the estimation error of $\mathbf{\Psi}$ is considered to decrease synergistically when the number of dimensions increases. 
Indeed, the estimation error decreased as the number of dimensions ranged from $D = 1$ to $D = 8$, whereas it increased as the number of dimensions ranged from $D = 8$ to $D = 16$.
This is probably because the calculation of the average absolute percentage error was performed using the Frobenius norm. Even if the estimation accuracy for each element in the matrix does not change, the overall error increases as the number of elements increases with the number of dimensions.

From the above, it is found that as the window length increases, the estimation of $\nu$ and $\mathbf{\Psi}$ is performed more accurately.
As the number of input dimensions increases, the estimation of these parameters is also performed more accurately; however, the estimation accuracy of $\mathbf{\Psi}$ decreases for higher-dimensional inputs.
Additionally, assuming the use of an international 10--20 electrode montage (i.e., the input signal has 19 dimensions), the model parameters can be estimated with an error rate of 10\% or less when the window length exceeds 5 s. 
In particular, setting the window length to 15 s can reduce the error rate to 5\%.

In Fig.~\ref{fig:ES_param}(c), the computation time tends to increase with the increasing window length $W$ and number of dimensions $D$, and the maximum average computation time was approximately 2 s for $W=100$ s and $D=19$.
However, even if all the channels in an international 10--20 electrode montage are used, it can be estimated in less than 0.5 s when $W$ is set to less than 20 s.
Therefore, the estimation results could be obtained almost in real-time if the sliding width $S$ of the window is set appropriately.
If the EEG is divided into multiple sub-bands using a filter bank, this calculation must be iterated for the number of sub-bands. 
However, since the parameter estimation in each band is completely independent, it can be estimated without loss of real-time performance by parallel computation using multiple threads.

The EEG analysis experiment showed that the proposed model is selected the greatest number of times in all frequency bands based on BIC (Table \ref{table:BIC}).
However, the ratio of the minimum BIC of the multivariate Gaussian distribution model was 17.92\% in the $\delta$ band, which was relatively higher than the others.
One possible explanation is that the distribution shape for each window length differs significantly in the low-frequency bands.
In addition, the number of parameters in the proposed model ($k = 191$) is slightly higher than that in the multivariate Gaussian distribution model ($k = 190$).
Accordingly, when the signal fits both the proposed and Gaussian models almost equally, the simpler Gaussian model shows better goodness-of-fit owing to the nature of BIC.
However, in the other frequency bands, the percentage at which the proposed model was selected based on the minimum BIC exceeds 95\%.
This is because the proposed model can change parameter $\nu$ to adapt to the shape of EEG distributions (i.e., Gaussianity) that change momentarily according to the state of brain activity.
By contrast, the Gaussian and Cauchy models have only a certain Gaussianity, and these two models are included in the proposed model as special cases (Gaussian: $\nu \rightarrow \infty$, Cauchy: $\nu = D$).
From the above, we can conclude that the proposed model is more suitable for EEG signals than the other models.
In addition, the ratio of the minimum BIC of the proposed model increases as the frequency band becomes higher.
This is because the EEG distribution changes depending on frequency characteristics, and those of high-frequency bands (at which signals change rapidly) are better fitted to the proposed model.

In Fig.~\ref{fig:Colormap}, the $1/\nu$ in the high-frequency band is particularly large in the epileptic seizure segments.
This can be confirmed from the results of all patients; therefore, the proposed method can quantitatively evaluate the change in non-Gaussianity during epileptic seizures as stochastic fluctuations.
However, at approximately 220 s in Fig. \ref{fig:Colormap}(a), a partial increase in $1/\nu$ was observed in intervals other than epileptic seizure segments.
This may have occurred because of artifacts in high-frequency bands, e.g., electromyograms caused by body movements.
The proposed feature, which characterizes stochastic fluctuations, may increase to some extent owing to non-stationary noise components superimposed onto the EEG.

In Fig. \ref{fig:dens}, the $1/\nu$ distributions of all patients in each frequency band show that the $1/\nu$ of non-epileptic seizure segments is distributed in a region smaller than 0.05 regardless of the frequency band.
By contrast, the $1/\nu$ of epileptic seizure segments is distributed more widely than that of non-epileptic seizure segments, and this tendency is most noticeable in the high-frequency $\gamma$ band.
This can also be confirmed from the effect size $g$, which is the standardized mean difference between the two groups.
It is considered that $1/\nu$ in the high-frequency band best reflects the characteristics of epileptic seizures because a large effect size was obtained in the $\gamma$ band.
This indicates that EEG signals in the high-frequency band exhibited a strong non-Gaussianity owing to epileptic seizures.
Previous studies reported that the activity in the $\gamma$ band, which is the high-frequency band of EEG signals, becomes intense during epileptic seizures~\cite{Kobayashi2004,Kobayashi2009,Benedek2016}.
The detailed mechanism of such $\gamma$ activity is still unknown; however, activities of inhibitory interneurons and electrical coupling through gap junctions were previously suggested~\cite{Park2012}.
These activities cause intermittent amplitude changes specific to epileptic seizures; consequently, the non-Gaussianity in the high-frequency band may be emphasized and $1/\nu$ increases.

For all features except ApEn, the maximum AUC was observed in the $\gamma$ band (Table.~\ref{table:AUC}).
In particular, the proposed $1/\nu$ in the $\gamma$ band showed the highest AUC compared to the other features and frequency bands, indicating that it has the highest classification ability for seizures and non-seizures.
This is also supported by the ROC curve and corresponding statistical test results shown in Fig.~\ref{fig:roc} and \ref{fig:roc_best}.
Although no significant difference was found between $1/\nu$ and RMS, their differences are highlighted from a different perspective in Fig.~\ref{fig:conf_mat}.
From the results of the confusion matrices, the conventional features, including RMS, tended to have high specificity (accuracy of detecting non-seizures) but relatively low sensitivity (accuracy of detecting seizures).
By contrast, the sensitivity of $1/\nu$ was higher than that of the conventional features, indicating that the proposed feature is superior in terms of accuracy in detecting epileptic seizures.
Furthermore, the diagonal elements of the confusion matrix of $1/\nu$ were equal, which means that the proposed feature in the $\gamma$ band can classify seizures and non-seizures in a balanced manner.
The confusion matrices having different characteristics in the proposed and conventional features, although with relatively close overall accuracy, suggests that these features reflect different aspects of EEG activity.
Thus, combining these features could result in higher classification performances.

These results revealed that the proposed feature defined by the stochastic EEG model has a relatively better seizure classification performance than the amplitude-based and amplitude-independent features that have been conventionally validated for epileptic seizure detection.

\section{Conclusion}
In this paper, EEG signals are modeled using a multivariate scale mixture model, allowing the representation of stochastic fluctuations in the variance-covariance matrix of EEGs.
We also proposed an EEG analysis method by combining the scale mixture model and a filter bank and introduced a feature $1/\nu$ characterizing the non-Gaussianity.
In this method, EEG signals are decomposed into several frequency bands, and time-series features for each frequency band were calculated based on the sliding window.

The simulation experiment evaluated the estimation accuracy of each parameter, which varied depending on the sample size and the number of dimensions. 
We demonstrated the proposed model to be most suitable in studying EEG signals that included epileptic seizures.
In addition, high accuracy (AUC = 0.881) in classifying seizure and non-seizure segments was obtained by focusing on the proposed feature $1/\nu$ in the $\gamma$ band.

This study focuses on the feature extraction part of the epileptic seizure detection problem.
Therefore, in addition to using a simple filter bank in the $\delta$--$\gamma$ bands for the frequency decomposition part, the classification of seizures/non-seizures was also performed by a simple threshold-based ROC analysis.
Applying the proposed method to automated detection systems for epileptic seizures requires searching for more effective EEG sub-bands and introducing machine learning-based classification techniques.
Furthermore, the proposed feature $1/\nu$ may be affected by artifacts such as electromyograms caused by body movement or stiffness owing to seizures.
Hence, an algorithm that detects and removes these artifacts will be introduced in the future.
We also plan to verify the effectiveness of the proposed method for epilepsy other than the focal seizures targeted in this study.

\appendix[Equivalent expression for multivariate scale mixture model]
This appendix shows the equivalent expressions of (\ref{eq:eq9}) and (\ref{eq:eq11}).
Equation (\ref{eq:eq11}) can be calculated as
\vspace{-2mm} 
\begin{align} 
	p(&\mathbf{x}_n) \notag \\
	&= \int \mathrm{IG}(\tau_n;\nu'/2,\nu'/2) \mathcal{N}(\mathbf{x}_n|\mathbf{0}, \tau_n \mathbf{\Psi}') \mathrm{d}{\tau_n} \nonumber\\
	&= \int \frac{\left(\frac{\nu'}{2}\right)^{\frac{\nu'}{2}}}{\Gamma \left(\frac{\nu'}{2}\right)} (\tau_n)^{-\frac{\nu'}{2}-1} \mathrm{exp} \left[-\frac{1}{\tau_n} \left(\frac{\nu'}{2} \right) \right] \nonumber\\
	&\quad\times\frac{1}{(2\pi)^{\frac{D}{2}}(\tau_n)^{\frac{D}{2}}|\mathbf{\Psi'}|^{\frac{1}{2}}} \mathrm{exp} \left[-\frac{1}{2}\mathbf{x}_n^\mathrm{T} (\tau_n\mathbf{\Psi'})^{-1} \mathbf{x}_n\right] \mathrm{d}{\tau_n} \nonumber\\
	&=\frac{1}{(2\pi)^{\frac{D}{2}}}\frac{\left(\frac{\nu'}{2}\right)^{\frac{\nu'}{2}}}{\Gamma \left(\frac{\nu'}{2}\right)}\frac{1}{|\mathbf{\Psi'}|^{\frac{1}{2}}} \nonumber\\
	&\quad\times \int (\tau_n)^{-\frac{\nu'+D}{2}-1} \mathrm{exp} \left[-\frac{1}{\tau_n} \left(\frac{\nu' + \Delta'}{2}\right) \right] \mathrm{d}{\tau_n} \nonumber \\
    &=\frac{1}{(2\pi)^{\frac{D}{2}}}\frac{\left(\frac{\nu'}{2}\right)^{\frac{\nu'}{2}}}{\Gamma \left(\frac{\nu'}{2}\right)}\frac{1}{|\mathbf{\Psi'}|^{\frac{1}{2}}} \frac{\Gamma \left(\frac{\nu'+D}{2}\right)}{\left(\frac{\nu'+\Delta'}{2}\right)^{\frac{\nu'+D}{2}}}\nonumber\\
	&\quad\times \int \frac{\left(\frac{\nu'+\Delta'}{2}\right)^{\frac{\nu'+D}{2}}}{\Gamma \left(\frac{\nu'+D}{2}\right)} (\tau_n)^{-\frac{\nu'+D}{2}-1} \nonumber\\
	&\quad\times \mathrm{exp} \left[-\frac{1}{\tau_n} \left(\frac{\nu' + \Delta'}{2}\right) \right] \mathrm{d}{\tau_n} \nonumber\\
	&=\frac{1}{(2\pi)^{\frac{D}{2}}}\frac{\left(\frac{\nu'}{2}\right)^{\frac{\nu'}{2}}}{\Gamma \left(\frac{\nu'}{2}\right)}\frac{1}{|\mathbf{\Psi'}|^{\frac{1}{2}}} \frac{\Gamma \left(\frac{\nu'+D}{2}\right)}{\left(\frac{\nu'+\Delta'}{2}\right)^{\frac{\nu'+D}{2}}}\nonumber\\
	&\quad\times \int \mathrm{IG} \left(\tau_n;\frac{\nu'+D}{2},\frac{\nu'+\Delta'}{2} \right) \mathrm{d}{\tau_n}.
	\label{eq:20}
\end{align}
Here, the integral of the probability density function over the entire space is equal to 1.
Hence, (\ref{eq:20}) can be expressed in the same form as (\ref{eq:eq9}).
\begin{align} 
	p(\mathbf{x}_n) &= \frac{\Gamma(\frac{\nu'+D}{2})}{\Gamma(\frac{\nu'}{2})} \frac{|{\bm \Psi'}|^{-\frac{1}{2}}}{\left(\pi \nu' \right)^{\frac{D}{2}}} \left(1+\frac{\Delta '}{\nu '} \right)^{-\frac{\nu'+D}{2}}.
\end{align}

\bibliographystyle{IEEEtran}
\bibliography{ref.bib}

\end{document}